\documentclass{Interspeech}
\usepackage{amsmath,comment}
\usepackage{
amsfonts,
mathrsfs,
url,
xurl,
cite,
booktabs,
tabularx,
array,
multirow,
threeparttable,
amssymb,
musicography,
}
\usepackage{xcolor}
\usepackage{arydshln}
\usepackage{svg}
\usepackage{multirow}
\usepackage{cite}

\interspeechcameraready 

\author[affiliation={}]{Daiki}{Takeuchi}
\author[affiliation={}]{Binh Thien}{Nguyen}
\author[affiliation={}]{Masahiro}{Yasuda}
\author[affiliation={}]{Yasunori}{Ohishi} 
\author[affiliation={}]{Daisuke}{Niizumi}
\author[affiliation={}]{Noboru~Harada}{}

\affiliation[nocounter]{}{NTT Corporation}{Japan}

\email{d.takeuchi@ntt.com}

\newcommand{\ci}[1]{\fontsize{6pt}{6pt}\selectfont{$\pm$ #1}}
\newcommand{\ar}[1]{{\fontsize{6pt}{6pt}\selectfont{#1}}}

\newcommand{\tabnote}[1]{\fontsize{9pt}{10pt}\selectfont{#1}}

\newcolumntype{C}[1]{>{\centering\arraybackslash}p{#1\tablelength}}
\newcolumntype{L}[1]{>{\arraybackslash}p{#1\tablelength}}

\newcommand{\mysubsec}[1]{\vspace{-2pt}\subsection{#1}\vspace{-2pt}}

\title{
CLAP-ART: 
Automated Audio Captioning\\ with Semantic-rich Audio Representation Tokenizer
}


\begin{document}
\maketitle

\begin{abstract}

Automated Audio Captioning (AAC) aims to describe the semantic contexts of general sounds, including acoustic events and scenes, by leveraging effective acoustic features. To enhance performance, an AAC method, EnCLAP, employed discrete tokens from EnCodec as an effective input for fine-tuning a language model BART. However, EnCodec is designed to reconstruct waveforms rather than capture the semantic contexts of general sounds, which AAC should describe. To address this issue, we propose CLAP-ART, an AAC method that utilizes ``semantic-rich and discrete'' tokens as input. CLAP-ART computes semantic-rich discrete tokens from pre-trained audio representations through vector quantization. We experimentally confirmed that CLAP-ART outperforms baseline EnCLAP on two AAC benchmarks, indicating that semantic-rich discrete tokens derived from semantically rich AR are beneficial for AAC.

% we propose CLAP-ART, a variant of EnCLAP, which introduces a new audio tokenizer called Audio Representation Tokenizer (ART) to replace EnCodec's discrete tokens with semantic-rich ones. 
% ART computes semantic-rich discrete tokens from pre-trained audio representations through vector quantization. 
% We confirmed that CLAP-ART outperforms EnCLAP on two AAC benchmarks, and ART provides semantic-rich discrete tokens that are effective for AAC.
% Comparison with state-of-the-art methods demonstrates that CLAP-ART's improvement in input features is comparable to that of introduced large language models.

\end{abstract}

\section{Introduction}
\label{sec:intro} 

%Automated Audio Captioning (AAC) is a multimodal text generation task that generates descriptions of general sounds~\cite{drossos2017automated, mei2022automated,xu2024beyond}. 
%AAC aims to understand and describe semantic contexts of general sounds, including acoustic events and scenes, unlike conventional audio classification tasks, which just classify labels.

Automated Audio Captioning (AAC) is a crossmodal text generation task focused on producing descriptions of diverse sounds~\cite{drossos2017automated, mei2022automated,xu2024beyond}. Unlike traditional audio classification tasks that assign labels to the sound, AAC seeks to comprehend and articulate the semantic contexts surrounding various acoustic events and scenes.

%In order to improve the quality of generated captions, recent studies have explored the use of pre-trained language models~\cite{kim2024enclap, gontier2021automated, mei2024wavcaps, haji2024taming,wu2024improving,koizumi2020audio, ghosh2024recap,liu2024enhancing}. 
%In these approaches, the language model receives the acoustic features as input to interpret the semantic contexts of input sounds.

To enhance the quality of generated captions, some studies have investigated the efficacy of pre-trained language models~\cite{koizumi2020audio, gontier2021automated, kim2024enclap,  mei2024wavcaps, wu2024improving,  ghosh2024recap, haji2024taming, liu2024enhancing}. These approaches involve feeding acoustic features into language models, which enables them to interpret the semantic contexts of the input sounds more effectively.

%Recent AAC methods utilize audio representation (AR) models to compute acoustic features that provide semantic information to the language model~\cite{kim2024enclap, mei2024wavcaps, haji2024taming,wu2024improving,ghosh2024recap,liu2024enhancing}.
%AR models are trained using various learning strategies.
%BEATs~\cite{chen2023beats} trains an audio encoder and an acoustic tokenizer using a combination of self-supervised learning of audio signals and supervised learning by labels in AudioSet~\cite{gemmeke2017audioset}.
%CLAP~\cite{wu2023large} uses cross-modal contrastive learning between an audio signal and the text caption. 
%AR features should capture semantic information because AR models have shown effectiveness in various downstream tasks related to general sounds. 
%In AAC methods, these semantic-rich AR features are typically fed to language models while remaining in a continuous feature representation.

Recent AAC methodologies employ Audio Representation (AR) models to derive acoustic features that convey semantic information to the language model~\cite{kim2024enclap, mei2024wavcaps, haji2024taming, wu2024improving, ghosh2024recap, liu2024enhancing}. AR models are trained using a variety of learning strategies. For instance, BEATs~\cite{chen2023beats} employs a combination of self-supervised learning of audio signals along with supervised learning using labels from AudioSet~\cite{gemmeke2017audioset}. Similarly, CLAP~\cite{wu2023large} leverages cross-modal contrastive learning to align audio signals with corresponding text captions. The ability of AR features to capture semantic information has been demonstrated across numerous downstream tasks associated with general sounds. In AAC frameworks, these semantically rich AR features are typically fed into language models in a continuous feature format.

% Among AAC methods leveraging AR features, EnCLAP~\cite{kim2024enclap} demonstrated providing acoustic features to the language model as a discrete token representation rather than a continuous feature representation improves AAC performance.  

%Among AAC methods that leverage AR features, EnCLAP~\cite{kim2024enclap} has demonstrated that AAC performance is improved by providing language models with acoustic features in discrete token representation rather than continuous feature representation.
%EnCLAP employs EnCodec~\cite{defossez2022high}, a neural audio codec, to compute discrete tokens and then inputs EnCodec discrete tokens along with CLAP audio embeddings into the language model BART~\cite{lewis2020bart}. 
%EnCLAP has verified that \textit{``pre-trained language models can leverage discrete inputs better compared to continuous input''}.
%This is likely because typical language models are pre-trained using discrete tokens of text.

Among the AAC methods that utilize AR features, EnCLAP~\cite{kim2024enclap} has shown that performance improves when acoustic features are provided to language models as discrete token representations instead of continuous formats. EnCLAP utilizes EnCodec~\cite{defossez2022high}, a neural audio codec, to generate discrete tokens, which are then fed into the BART language model~\cite{lewis2020bart} alongside CLAP audio embeddings. This approach has validated the assertion that ``pre-trained language models can leverage discrete inputs better compared to their continuous inputs.'' This performance enhancement is likely attributable to the fact that standard language models are trained on discrete textual tokens.

%However, EnCodec discrete tokens are suboptimal for AAC because EnCodec is designed to encode audio waveforms into discrete tokens so that the original waveform can be decoded precisely, rather than generating discrete tokens that capture the semantic information in sounds, which AAC should describe. 
%Indeed, EnCodec discrete tokens have worse performance in various downstream tasks related to general sounds than AR features~\cite{liu2024semanticodec}.

However, it should be noted that EnCodec discrete tokens may not be optimal for AAC, as EnCodec is primarily designed to compress audio waveforms into discrete tokens for precise reconstruction of the original waveform rather than producing tokens that encapsulate the semantic richness of the sounds, which AAC aims to describe. 
It has been found that EnCodec discrete tokens yield inferior performance in various downstream tasks related to general sounds when compared to AR features~\cite{liu2024semanticodec}.

%Based on the above, we hypothesize that semantic-rich discrete token representation improves AAC performance.
%Semantic-rich discrete tokens should reflect the semantic information of general audio, incorporating information from other modalities, such as labels of acoustic events and scenes.
%BEATs' acoustic tokenizer is trained by distilling knowledge from the BEATs' audio encoder and can generate such semantic-rich discrete tokens.
%In contrast, EnCodec focuses on waveform compression and reconstruction, meaning its discrete tokens would not be semantic-rich. 
%SpeechTokenizer~\cite{zhang2024speechtokenizer} generates discrete tokens reflecting the knowledge of HuBERT~\cite{hsu2021hubert}, a self-supervised speech representation model. 
%However, since HuBERT's knowledge is derived from phonemes and linguistic semantics of spoken words, SpeechTokenizer's discrete tokens are also not semantically rich for general sounds.

Based on the preceding analysis, we hypothesize that ``semantic-rich and discrete'' token representations can enhance the performance of AAC. These tokens should encapsulate the semantic information intrinsic to general audio, integrating insights from various modalities, such as labels associated with acoustic events and scenes. The acoustic tokenizer utilized in BEATs distills knowledge from BEATs' audio encoder, enabling it to generate discrete tokens that embody this semantic richness.
In contrast, EnCodec primarily focuses on compression and reconstruction, which limits its discrete tokens' ability to capture semantic richness. Similarly, the SpeechTokenizer~\cite{zhang2024speechtokenizer} produces discrete tokens that derive their information from HuBERT~\cite{hsu2021hubert}, a self-supervised model focused on speech representation. However, because HuBERT's representation is anchored in phonemes and the linguistic semantics of the spoken language, the discrete tokens generated by the SpeechTokenizer lack the semantic richness necessary for representing general sounds effectively.

\begin{figure}[t]
    \centering
    \includegraphics[width=0.85\columnwidth]{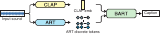} 
    \vspace{-5pt}
    \caption{
    Overview of CLAP-ART.
    CLAP-ART utilizes ART semantic-rich discrete tokens as BART input along with a CLAP audio embedding.
    }
    \label{fig:clapart_overview}
    \vspace{-15pt}
\end{figure}

\begin{figure*}[t]
    \centering
    \includegraphics[width=1.65\columnwidth]{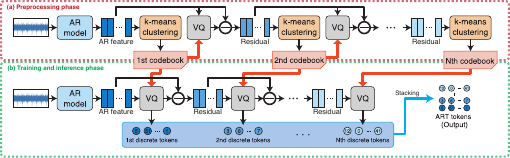} 
    \vspace{-5pt}
    \caption{
    Details of BEATs-RVQ. 
    The preprocessing phase (a) obtains codebooks in advance of fine-tuning BART.
    The training and inference phase (b) converts input sound to discrete tokens.
    }
    \label{fig:beats-rvq_details}
     \vspace{-15pt}
\end{figure*}

In this paper, we propose CLAP-ART, an AAC method that utilizes semantic-rich discrete tokens as input features for a language model. 
To generate these tokens, we introduce an audio representation tokenizer (ART), which extracts discrete tokens from AR features.
We implement two types of ART: (1) BEATs-AT, which naively employs BEATs' acoustic tokenizer, and (2) BEATs-RVQ, which quantizes BEATs' audio encoder output by applying k-means clustering in the manner of  Residual Vector Quantization~(RVQ)~\cite{defossez2022high,zeghidour2021soundstream}. 
BEATs-AT generates a single-layer discrete token, whereas BEATs-RVQ produces multi-layer discrete tokens, which are expected to enhance AAC performance by providing richer semantic information. 
CLAP-ART feeds these semantic-rich discrete tokens alongside a CLAP audio embedding into BART to generate captions as illustrated in Fig.~\ref{fig:clapart_overview}.

In the experiments to validate our hypothesis, we confirmed that CLAP-ART improves AAC performance. 
CLAP-ART outperformed the baseline EnCLAP on two AAC benchmarks. 
We also confirmed that SpeechTokenizer's discrete tokens did not improve AAC performance, indicating that they are unsuitable for AAC. 
Furthermore, BEATs-RVQ, which enables multi-layer discrete tokens, outperformed BEATs-AT, verifying that multi-layer discrete tokens can capture more semantic information effectively for AAC.
These experiments demonstrate that semantic-rich discrete tokens derived from semantically rich AR are beneficial for AAC.

\section{Related work}
\mysubsec{Automated Audio Captioning Methods using Pre-trained Language Model}
Pre-trained language models have been utilized in some studies to improve AAC performance~\cite{
kim2024enclap, 
gontier2021automated, 
mei2024wavcaps, 
haji2024taming,
wu2024improving,
koizumi2020audio, 
ghosh2024recap,
liu2024enhancing
}.
In order to provide semantic information to language models, pre-trained AR features are used as input.
Typical approaches simply pass AR features to language models through a linear layer~\cite{mei2024wavcaps}, Conformer layer~\cite{wu2024improving}, or Querying Transformer layer~\cite{haji2024taming, liu2024enhancing}. 
BART-tags~\cite{gontier2021automated} provides an AR feature and the BART embedding of the labels predicted by the audio tagging model to the language model BART. 
RECAP~\cite{ghosh2024recap} uses captions of similar sounds to an input sound with an AR feature, allowing it to adapt to out-of-domain acoustic events and scenes.
Koizumi et al.~\cite{koizumi2020audio} uses a similar procedure as RECAP to tackle the difficulty of adapting a powerful pre-trained language model to the AAC task.

While AR features are effective for the AAC task, there is still room to consider in the format of acoustic features provided to language models. 
EnCLAP demonstrated that discrete token representation is more effective for fine-tuning language models than continuous feature representation.
The details of EnCLAP are shown in Section~\ref{sec:enclap}.

\mysubsec{Audio Discrete Token with Semantic Information}

In recent years, several methods have been tried to generate discrete tokens for general sounds.
Neural audio codecs, such as Encodec~\cite{defossez2022high}, utilize deep neural networks for compressing an audio waveform into a discrete token representation and reconstructing the original waveform from these discrete tokens.
SemantiCodec~\cite{liu2024semanticodec} achieves a lower bitrate codec compared with those of other methods by leveraging pre-trained AR features. 
% It also demonstrates superior performance in various audio classification tasks compared to other neural audio codecs. 
% However, the performance of these audio classification tasks is inferior to the original AR features employed in SemantiCodec.
BEATs \cite{chen2023beats} incorporates an acoustic tokenizer in addition to the audio encoder.
This tokenizer outputs single-layer discrete tokens.
Since BEATs' acoustic tokenizer is designed to represent semantic information, it has achieved a level of performance comparable to that of BEATs' audio encoder in environmental sound classification tasks.

% BEATs参考: "the tokenized discrete labels are optimized to contain more semantic-rich knowledge from the teacher and less redundant information of the input audio."

Some studies in speech processing have explored the use of discrete token representations derived from the features of self-supervised speech models and examined their applications~\cite{Baevski2020vq-wav2vec, mousavi2024how, zhang2024speechtokenizer}. 
SpeechTokenizer leverages HuBERT's knowledge to learn unified speech discrete tokens that encapsulate both phonetic and semantic information. 
The effectiveness of these discrete tokens from SpeechTokenizer has been demonstrated in speech-related tasks such as speech recognition, zero-shot text-to-speech, and one-shot voice conversion. 
However, the phonetic and semantic information of speech is specific to spoken language and differs from the semantic context of general sounds, which is the focus of AAC. 
We used SpeechTokenizer as a comparison method to validate the difference between speech semantics and general sound semantics.

\section{Baseline: EnCLAP}
\label{sec:enclap}
EnCLAP leverages EnCodec alongside the CLAP audio encoder to extract input features for BART fine-tuning. 
First, EnCLAP transforms an input audio waveform into a CLAP audio embedding and EnCodec discrete tokens.
The CLAP audio embedding is projected onto the BART input embedding space by using a linear layer, resulting in $e_{\rm clap}\in\mathbb{R}^{D_{\rm b}}$, where ${D_{\rm b}}$ is the hidden size of BART.
EnCodec discrete tokens are converted into embeddings by using embedding layers corresponding to the respective layers of discrete tokens and then summed to form a sequence $e_{\rm encodec}\in\mathbb{R}^{L \times D_{\rm b}}$, where $L$ is the time length of the discrete tokens.
Next, $e_{\rm encodec}$ is enclosed with BART's special tokens $\texttt{<bos>}$ and $\texttt{<eos>}$
And then, positional embeddings $e_{\rm pos}$ is applied to it:
\begin{equation}
\label{eq:enclap_seq_fet}
I_{\rm seq} = \left[
e_{\rm bos}, 
e_{\rm encodec}, 
e_{\rm eos}
\right] + e_{\rm pos} \in \mathbb{R}^{(L+2) \times D_{\rm b}}.
\end{equation}
Finally, $e_{\rm clap}$ and $I_{\rm seq}$ are concatenated to produce input for BART:
\begin{equation}
I_{\rm EnCLAP} = \left[
e_{\rm clap},
I_{\rm seq}
\right] \in \mathbb{R}^{(L+3) \times D_{\rm b}}.
\end{equation}

EnCLAP is trained using a combination of the cross-entropy loss $\mathcal{L}_{\rm caption}$ for caption generation and the masked codec modeling (MCM) loss $\mathcal{L}_{\rm mcm}$. The total loss is given by $\mathcal{L}_{\rm total} = \mathcal{L}_{\rm caption} + \lambda \mathcal{L}_{\rm mcm}$, where $\lambda$ is a weighting parameter.
% 詳細はEnCLAPの論文を読んでね的な一文を入れる？

\section{Proposed Method}
\label{sec:proposal}
CLAP-ART leverages the audio representation tokenizer~(ART) proposed in this paper alongside the CLAP audio encoder to extract input features for BART fine-tuning.

\mysubsec{Audio Representation Tokenizer}
ART converts input sounds into semantic-rich discrete tokens based on the knowledge of AR models.
We implemented two types of ART: BEATs-AT and BEATs-RVQ. 
BEATs-AT naively uses BEATs’ acoustic tokenizer.
% to generate single-layer discrete tokens.
BEATs-RVQ is a newly designed ART, which quantizes continuous features output from BEATs’ audio encoder by RVQ~\cite{zeghidour2021soundstream,defossez2022high}.

The details of BEATs-RVQ are illustrated in Fig.~\ref{fig:beats-rvq_details}.
Prior to the training, we create the RVQ codebooks on the training dataset by using k-means clustering of the continuous audio features of audio samples encoded by the AR model.
% Although BEATs’ audio encoder extracts features from flattened time-frequency patches, BEATs-RVQ directly uses the sequence of flattened features for RVQ without reshaping it to reflect the original time-frequency structure.
Although the BEATs' audio encoder outputs $768\times FT$ features from $F \times T$ (frequency $\times$ time) patches, BEATs-RVQ feeds them directly into RVQ as a sequence of 768-dimensional vectors (length $FT$) without reshaping them to reflect the original time-frequency structure (e.g., $768F \times T$).
In the training and inference stage, the AR model encodes the input into features, and then the RVQ layer vector-quantizes the features into discrete tokens.
The VQ procedure follows the original RVQ steps in \cite{zeghidour2021soundstream,defossez2022high}: 
Project the input vector (feature) onto the closest entry in the first codebook, compute the residual after quantization, quantize it using the next codebook, and repeat the last two steps until the last codebook.
BEATs-RVQ can generate multi-layer discrete tokens, which are expected to improve AAC performance by packing more semantic information than in BEATs-AT discrete tokens, which are only single-layer.

\mysubsec{Training with ART Discrete Tokens}
We trained CLAP-ART on the EnCLAP framework.
CLAP-ART calculates embeddings $e_{\rm art}\in\mathbb{R}^{L_ \times D_{\rm b}}$ from ART discrete tokens by using the same procedure for embedding EnCodec discrete tokens.
$e_{\rm art}$ is used instead of $e_{\rm encodec}$ as follows:
\begin{align}
    I_{\rm seq} &= [
    e_{\rm bos}, 
    e_{\rm art}, 
    e_{\rm eos}
    ] + e_{\rm pos}
    \in \mathbb{R}^{(L+2) \times D_{\rm b}}, \\
    I_{\rm CLAP\text{-}ART} &=
    [
    e_{\rm clap},
    I_{\rm seq}
    ]
    \in \mathbb{R}^{(L+3) \times D_{\rm b}}.
\end{align}
CLAP-ART uses $I_{\rm CLAP\text{-}ART}$ instead of $I_{\rm EnCLAP}$ and follows the EnCLAP's training procedure as mentioned in Section~\ref{sec:enclap}.

\begin{table*}[t!]
\caption{
Comparison of CLAP-ART with conventional methods on AudioCaps and Clotho.
Bold indicates the best score for each metric.
}
\vspace{-8pt}
\label{tab:main_results}
\centering
\resizebox{1.7\columnwidth}{!}{
\begin{tabular}{l|ccccc|ccccc} 
\toprule 
% \hline \hline
& \multicolumn{5}{c}{AudioCaps} & \multicolumn{5}{c}{Clotho (Pre-trained on AudioCaps)}  \\ 
\cmidrule(lr){2-6} \cmidrule(lr){7-11}
Method & METEOR & CIDEr & SPICE & SPIDEr & FENSE & METEOR & CIDEr & SPICE & SPIDEr & FENSE \\ \midrule
% \textbf{Baseline} &&&&&&\\
\textbf{Conventional} &&&&&&\\
% EnCLAP-base~\cite{kim2024enclap} 
EnCLAP~\cite{kim2024enclap}~\ar{(CLAP-EnCodec)}
& 24.2~\ci{0.36} & 74.4~\ci{1.43} & 18.1~\ci{0.21} & 46.3~\ci{0.76} & 64.2~\ci{0.47}
& 18.2~\ci{0.11} & 45.4~\ci{0.69} & 12.8~\ci{0.20} & 29.1~\ci{0.38} & 50.6~\ci{0.19} 
\\
CLAP-SpeechTokenizer 
& 24.3~\ci{0.08} & 74.7~\ci{1.15} & 17.8~\ci{0.45} & 46.3~\ci{0.76} & 64.2~\ci{0.41}
& 18.2~\ci{0.15} & 45.3~\ci{0.70} & 12.9~\ci{0.15} & 29.1~\ci{0.42} & 50.1~\ci{0.28} \\

CLAP-HuBERT-RVQ 
& 24.5~\ci{0.42} & 74.6~\ci{1.59} & 18.0~\ci{0.16} & 46.3~\ci{0.83} & 64.6~\ci{0.33}
& 18.2~\ci{0.15} & 45.8~\ci{0.72} & 13.1~\ci{0.13} & 29.4~\ci{0.39} & 50.4~\ci{0.28} \\
% CLAP-Wav2Vec2 \\
% EnCLAP-large~\cite{kim2024enclap} & 24.8~\ci{0.22} & {25.5} & 75.3~\ci{2.50} & {80.3} & 18.4~\ci{0.35} & {18.8} & 46.9~\ci{1.41} & {49.5} \\
\midrule
% \hline
\textbf{Proposed} &&&&&&\\

CLAP-ART(BEATs-AT) 
& 25.1~\ci{0.45} & 78.0~\ci{1.30} & 18.4~\ci{0.25} & 48.2~\ci{0.74} & 64.9~\ci{0.32}
& 18.6~\ci{0.11} & 47.3~\ci{0.48} & 13.2~\ci{0.10} & 30.3~\ci{0.20} & 50.8~\ci{0.13} \\
CLAP-ART(BEATs-RVQ)
& \textbf{25.6}~\ci{0.39} & \textbf{80.7}~\ci{1.22} & \textbf{18.8}~\ci{0.25} & \textbf{49.8}~\ci{0.61} & \textbf{65.5}~\ci{0.11} 
& \textbf{18.7}~\ci{0.16} & \textbf{47.5}~\ci{1.16} & \textbf{13.3}~\ci{0.16} & \textbf{30.4}~\ci{0.65} & \textbf{51.1}~\ci{0.55} 
\\
% CLAP-ART~\ar{(M2D-CLAP)}
% & 24.7~\ci{0.19} & 76.4~\ci{1.40} & 18.0~\ci{0.18} & 47.2~\ci{0.77} & 65.2~\ci{0.12}  
% & 18.4~\ci{0.12} & 46.7~\ci{0.62} & 12.9~\ci{0.12} & 29.8~\ci{0.37} & 50.7~\ci{0.25} 
% \\

% % CLAP-ART~\ar{(M2D2AS+)}
% & 24.8~\ci{0.20} & 77.7~\ci{1.07} & 18.1~\ci{0.32} & 47.9~\ci{0.57} &
% & 18.0~\ci{0.18} & 44.5~\ci{0.96} & 12.9~\ci{0.22} & 28.7~\ci{0.56} &
% \\
\bottomrule
% \multicolumn{9}{l}{\tabnote{$^{*}$ 
% Claimed is either a) or b). 
% }}\\ \addlinespace[-0.1cm]
% \multicolumn{9}{l}{\tabnote{
% \, a) The results claimed in the corresponding papers, denoted by ${\Lsh}$
% }}\\ \addlinespace[-0.1cm]
% \multicolumn{9}{l}{\tabnote{
% \, b) Scores of the system that performed the best SPIDEr among systems trained with 6 different random seeds, denoted by ${\dagger}$
% }}
\end{tabular}
}
\vspace{-10pt}
\end{table*}

\begin{table}[t!]
\caption{
Ablation study of CLAP audio embedding on AudioCaps to verify semantic-richness of ART discrete tokens.
}
\vspace{-8pt}
\label{tab:abblation_clap}
\centering
\resizebox{0.95\columnwidth}{!}{%
\begin{tabular}{l|cccccccccc} 
\toprule
& \multicolumn{2}{c}{METEOR} & \multicolumn{2}{c}{CIDEr} & \multicolumn{2}{c}{SPICE}  &  \multicolumn{2}{c}{SPIDEr} & \multicolumn{2}{c}{FENSE}\\ 
\cmidrule(lr){2-3} \cmidrule(lr){4-5} \cmidrule(lr){6-7} \cmidrule(lr){8-9} \cmidrule(lr){10-11}
Method & Mean & Diff & Mean & Diff & Mean & Diff & Mean & Diff & Mean & Diff  \\
\midrule
EnCLAP-base~\cite{kim2024enclap} & 24.2 & --   & 74.4 & --    & 18.2 &  --  & 46.3 & --    & 64.2 & --\\
\, w/o CLAP$^{\musFlat}$         & 18.9 & -5.3 & 48.4 & -26.0 & 13.0 & -5.2 & 30.7 & -15.6 & 49.7 & -14.5\\
\midrule
CLAP-ART(BEATs-RVQ)
% ~\ar{(BEATs)}    
& 25.6 & --   & 80.7 & --   & 18.8 & --   & 49.8 & --   & 65.5 & --\\
\, w/o CLAP$^{\musFlat}$   & 25.0 & -0.6 & 77.1 & -3.6 & 18.1 & -0.7 & 47.6 & -2.2 & 63.9 & -1.6\\
\bottomrule
% \addlinespace[0.01cm]
\multicolumn{9}{l}{\tabnote{$^{\musFlat}$ CLAP audio embedding is disabled. That is, $I_{\rm seq}$ is used for $I_{\rm EnCLAP}$ or $I_{\rm CLAP\text{-}ART}$}.}\\
\end{tabular}
}
\vspace{-5pt}
\end{table}

\begin{table}[t!]
\caption{
Comparison with variation of the number of RVQ layers in BEATs-RVQ on AudioCaps.
$N$~indicates the number of RVQ layers.
}
\vspace{-8pt}
\label{tab:abblation_rvq}
\centering
\resizebox{0.85\columnwidth}{!}{%
\begin{tabular}{lc|cccccc} 
\toprule
% & \multicolumn{2}{c}{VQ setup} \\ \cmidrule(lr){2-3}
Method & $N$ 
% & Size 
& METEOR & CIDEr & SPICE & SPIDEr &FENSE \\
\midrule
CLAP-ART(BEATs-RVQ)
& 16  & \textbf{25.6} & \textbf{80.7} & \textbf{18.8} & \textbf{49.8} & \textbf{65.5}\\
& 8  & 25.5 & 80.6 & 18.8 & 49.7 & 65.2 \\
& 4  & 25.3 & 78.6 & 18.6 & 48.6 & 65.0 \\
& 2  & 25.2 & 79.2 & 18.4 & 48.8 & 65.1 \\
& 1  & 25.0 & 78.6 & 18.4 & 48.5 & 65.0 \\ 
% \cmidrule(lr){2-8}
% & 1 & 32768 & 24.2 & 75.0 & 17.8 & 46.4 & 64.6 \\
% & 1 & 4096 & 25.3 & 79.1 & 18.5 & 48.8 & 65.0 \\
\midrule
CLAP-ART(BEATs-AT)      & 1  & 25.1 & 78.0 & 18.4 & 48.2 & 64.9\\
% \midrule
% SemantiCodec \ar{($\mathcal{V}=32768$)}% ~\ar{(100~tokens/sec, $N_$)}
% % SemantiCodec
% & 1 & 32768 & 23.8 & 73.1 & 17.6 & 45.3 & 63.8\\
% SemantiCodec \ar{($\mathcal{V}=4096$)}% ~\ar{(100~tokens/sec, $N_$)}
% & 1 & 4096 & 24.1 & 74.9 & 17.9 & 46.4 & 64.4 \\

% \midrule\midrule

% \multicolumn{7}{l}{\textbf{Without attention to General sounds' semantics}}\\
% EnCodec~\ar{(EnCLAP)}~\cite{kim2024enclap} & 16  & 24.2 & 74.4 & 18.1 & 46.3 & 64.2\\
% % & 1 & \\
% \midrule
% SpeechTokenizer~\cite{zhang2024speechtokenizer} & 8  & 24.2 & 74.1 & 18.0 & 46.0  & 64.3 \\
% & 1  & 24.0 & 72.8 & 17.7 & 45.2  & 63.9 \\
\bottomrule
\end{tabular}
}
\vspace{-5pt}
\end{table}

\begin{table}[t!]
\caption{
% Comparison with conventional methods on AudioCaps.
Comparison with SOTA methods on AudioCaps.
\#LM Params. indicates the number of parameters of the pre-trained language model.
}
\vspace{-8pt}
\label{tab:audiocaps_compare_sota}
\centering
\resizebox{0.98\columnwidth}{!}{
\setlength{\tabcolsep}{4pt}
\begin{tabular}{lcccccc} 
\toprule
Method & \#LM Params. & METEOR & CIDEr & SPICE & SPIDEr & FENSE \\
\midrule
\multicolumn{6}{l}{\textbf{Reported in their papers}} \\
BART-tags~\cite{gontier2021automated}
& 406M & 24.1  & 75.3  & 17.6  & 46.5 & --\\
% AL-MixGen~\cite{kim2022exploring} 
%  & 24.2  & 76.9  & 18.1  & 47.5 & --\\
HTSAT-BART$^{+}$~\cite{mei2024wavcaps}
& 139M & 25.0  & 78.7  & 18.2  & 48.5 & -- \\
% CNeXt-trans~\cite{labb2024conette} 
% & 9M & 25.2  & 80.6  & 18.4  & 49.5 & 64.3 \\
AutoCap (w/o metadata input)~\cite{haji2024taming} 
& 139M & 24.6  & 77.3  & 18.2  & 47.8 & -- \\
AutoCap (w/o metadata input)$^{+}$~\cite{haji2024taming}
& 139M & 25.6  & 80.4  & 19.0  & 49.7 & -- \\
EnCLAP-base$^{\ast}$~\cite{kim2024enclap}
& 139M & 24.7 & 78.0 & 18.6 & 48.3 & -- \\
EnCLAP-large$^{\ast}$~\cite{kim2024enclap}
& 406M & 25.5 & 80.3 & 18.8 & 49.5 & -- \\
LOAE$^{+}$\cite{liu2024enhancing} 
& 7B & 26.7 & 81.6 & 19.3 & 50.5 & 66.2 \\
% EnCLAP++-base
% & 25.7 & 81.5 & 18.8 & 50.1 & 66.1 \\
% EnCLAP++-large
% & 26.9 & 82.3 & 19.7 & 51.0 & 66.5 \\
% SLAM-AAC
% & 26.8 & 84.1 & 19.4 & 51.8 & 66.8 \\
\midrule
\multicolumn{6}{l}{\textbf{Ours (Averaged scores)}} \\
CLAP-ART(BEATs-RVQ)-base$^{\ast\dagger}$
& 139M & 25.6 & 80.7 & 18.8 & 49.8 & 65.5 \\
CLAP-ART(BEATs-RVQ)-large$^{\ast\dagger}$
& 406M & 25.7 & 80.8 & 19.1 & 50.0 & 65.4 \\
\\ \addlinespace[-9pt]

\multicolumn{6}{l}{\textbf{The best performing system results of ours (Reference)}} \\
{\color{gray} CLAP-ART(BEATs-RVQ)-base$^{\ast\ddagger}$ }
& {\color{gray} 139M }& {\color{gray} 25.8} & {\color{gray} 82.0} & {\color{gray} 19.0} & {\color{gray} 50.5} & {\color{gray} 65.4} \\
{\color{gray} CLAP-ART(BEATs-RVQ)-large$^{\ast\ddagger}$ }
& {\color{gray} 406M }& {\color{gray} 25.9} & {\color{gray} 81.5} & {\color{gray} 19.6} & {\color{gray} 50.6} & {\color{gray} 65.2} \\
\bottomrule
\\ \addlinespace[-0.3cm]
\multicolumn{7}{l}{
\tabnote{$^{+}$Methods used pre-training with extra audio-text paired datasets such as WavCaps.
% Training of our CLAP-ART used only AudioCaps.
}}\\ \addlinespace[-0.05cm]

\multicolumn{6}{l}{
\tabnote{$^{\ast}$Only EnCLAP and CLAP-ART utilized discrete tokens as input features.
}}\\ \addlinespace[-0.05cm]

\multicolumn{6}{l}{
\tabnote{$^{\dagger}$Average scores in six experiments.
}}\\ 
\addlinespace[-0.05cm]

\multicolumn{6}{l}{\tabnote{$^{\ddagger}$The scores obtained by the system with the best spider score in six experiments.
}}\\ 
\addlinespace[-0.05cm]
% \multicolumn{9}{l}{\tabnote{
% \, b) Scores of the system that performed the best SPIDEr among systems trained with 6 different random seeds, denoted by ${\dagger}$
% }}
\end{tabular}
}
\vspace{-5mm}
\end{table}

\section{Experiments}
\mysubsec{Experimental Setup}\label{sec:exp-setup}
We conducted experiments on two AAC datasets: AudioCaps~\cite{kim2019audiocaps} and Clotho~\cite{drossos2020clotho}.
% In the experiments on AudioCaps, we conducted the training and evaluation using only AudioCaps.
% In the experiments on Clotho, we first conducted pre-training using AudioCaps and then the training and evaluation using Clotho.
For the experiments on AudioCaps, we trained and evaluated models solely on  AudioCaps.
For the experiments on Clotho, we first pre-trained models on AudioCaps and then fine-tuned and evaluated them on Clotho.

In the setup of CLAP-ART, we utilized the acoustic tokenizer and the audio encoder of BEATs~\ar{iter3+} for BEATs-AT and BEATs-RVQ, respectively.
For BEATs-RVQ, the number of RVQ layers and codebook size were 16 and 1024, respectively, which were the same as those for EnCodec in EnCLAP.
We used BART as the pre-trained language model to generate captions.
Basically, the base model of BART was used. 
The experiments shown in Table~\ref{tab:audiocaps_compare_sota} used both the base and large models of BART.

We used the same training and inference settings with EnCLAP for the sake of simple comparison, except for the learning rate and pre-training setup for Clotho, which we set based on grid search.
The number of epochs was 15. 
The loss weighting parameter $\lambda$ was fixed to $0.7$.
15 \% of the input tokens were masked for MCM. 
The mask length was 10. 
The optimizer was AdamW~\cite{loshchilov2017decoupled} with $\beta_1 = 0.9$ and $\beta_2 = 0.999$. 
Weight decay was set to 0.01.
Label smoothing where a factor was 0.2 was used for loss for only text generation $\mathcal{L}_{\rm caption}$. 
We used the inverse square root learning rate scheduler with a linear warm-up.
The experiment on AudioCaps used a peak learning rate of 6e-5 and 2000 warm-up steps,
In the experiment on Clotho, we conducted pre-training on AudioCaps in 5 epochs and then trained on Clotho with a peak learning rate of 2e-4 and no warmup.
The inference used a beam search with a beam size of 4.
We implemented the above in accordance with the official implementation of EnCLAP\footnote{\url{https://github.com/jaeyeonkim99/EnCLAP}} to reproduce the baseline.

To ensure that the results of this research are reproducible, we specifically experimented with EnCLAP, CLAP-ART, and other variations using six different random seeds and calculated statistics as the final results.
For the evaluation metrics, we computed METEOR~\cite{denkowski2014meteor}, CIDEr~\cite{vedantam2015cider}, SPICE~\cite{anderson2016spice}, SPIDEr~\cite{liu2017improved}, and FENSE~\cite{zhou2022can} by using the \textit{aac-metrics} library\footnote{\url{https://github.com/Labbeti/aac-metrics}}.

\mysubsec{Comparison with baseline and our CLAP-ART}
We found that CLAP-ART, supported by ART discrete tokens, improves AAC performance by comparing it with EnCLAP. The results are shown in Table~\ref{tab:main_results}.
EnCLAP is the baseline of this experiment.
CLAP-SpeechTokenizer used SpeechTokenizer's discrete tokens instead of ART ones.
CLAP-HuBERT-RVQ converted HuBERT features into discrete tokens by using the same procedure as BEATs-RVQ.
CLAP-ART(BEATs-AT) and CLAP-ART(BEATs-RVQ) utilized BEATs-AT and BEATs-RVQ as ART, respectively.
Each score denotes the average score and sample standard deviation of six experiments with different random seeds.

From these results, we confirmed that CLAP-ART outperforms EnCLAP across all evaluation metrics and benchmarks, 
supporting our hypothesis that semantic-rich discrete tokens are beneficial to enhancing AAC performance. 
In particular, both CLAP-ART(BEATs-AT) and CLAP-ART(BEATs-RVQ) demonstrated more significant improvements on AudioCaps.
For instance, CLAP-ART(BEATs-RVQ) improved the SPIDEr score to $+3.5$, while that on Clotho was $+1.3$.
Since the BEATs$_{\rm iter3+}$ was pre-trained using the AudioSet supervision, it is reasonable for it to be effective for AudioCaps, a subset of AudioSet.
CLAP-SpeechTokenizer and CLAP-HuBERT-RVQ did not improve AAC performance from the level of EnCLAP, indicating that knowledge of speech phonemes and linguistic semantics is not beneficial for AAC.

\mysubsec{Ablation study of CLAP audio embedding}
To validate the semantic richness of discrete tokens generated by ART, we conducted an ablation experiment on the CLAP audio embedding on AudioCaps.
We compared the performances of EnCLAP and CLAP-ART(BEATs-RVQ) with and without the CLAP embedding input enabled. When disabled, the AAC system relied solely on the semantic information in the discrete tokens of the EnCodec for EnCLAP or those of BEATs-RVQ for CLAP-ART.

Table~\ref{tab:abblation_clap} shows the average scores and the difference between with and without the CLAP embedding. The columns of ``Diff'' clearly show that degradation in CLAP-ART(BEATs-RVQ)'s scores was smaller than the EnCLAP's (compare, e.g.,  CLAP-ART(BEATs-RVQ)'s $-2.2$ vs. EnCLAP's $-15.6$ in SPIDEr).
% such as CLAP-ART(BEATs-RVQ)'s $-2.2$ vs. EnCLAP's $-15.6$ in SPIDEr. 
CLAP-ART(BEATs-RVQ) w/o CLAP also achieved comparable scores to EnCLAP using CLAP~($47.6$ vs. $46.3$ in SPIDEr). 
These results demonstrated that ART discrete tokens effectively capture semantic information of general sounds, improving AAC performance.

\mysubsec{Verification that semantic-rich multi-layer discrete tokens from BEATs-RVQ capture more semantic information}
This experiment investigated whether BEATs-RVQ improves AAC performance by capturing more semantic information through multi-layer codebooks.
In addition to the results shown in Table~\ref{tab:main_results}, we further evaluated several variants of CLAP-ART (BEATs-RVQ) with different numbers of RVQ layers.

The results are shown in Table~\ref{tab:abblation_rvq}.
CLAP-ART (BEATs-AT) and CLAP-ART (BEATs-RVQ) with $N=1$ showed mostly equivalent performance.
CLAP-ART (BEATs-RVQ) can leverage RVQ to increase the number of layers to $N=16$, achieving higher scores than CLAP-ART (BEATs-AT).
This demonstrates that multi-layer discrete tokens from BEATs-RVQ improve AAC performance by packing more semantic information.

\mysubsec{Comparison with state-of-the-art AAC methods}
We conducted a comparison with conventional AAC methods. 
As shown in Table~\ref{tab:audiocaps_compare_sota}, the results from conventional methods are based on their original papers.
Among these, HTSAT-BART~\cite{mei2024wavcaps} and LOAE~\cite{liu2024enhancing} were trained with extra audio-text paired datasets such as WavCaps~\cite{mei2024wavcaps} (denoted by $^+$), whereas BART-tags~\cite{gontier2021automated}, EnCLAP~\cite{kim2024enclap}, and our CLAP-ART were trained using only AudioCaps.
AutoCap~\cite{haji2024taming} reported the results for both versions: one trained solely on AudioCaps and another utilizing extra datasets.
In this experiment, we used only audio input for both versions of AutoCap to compare with other methods.
Although we focused on audio-only input, AutoCap has the potential to improve caption quality by incorporating textual metadata as an additional input.
As for our CLAP-ART(BEATs-RVQ), we compared two systems: one using BART's base model (CLAP-ART(BEATs-RVQ)-base) and the other using BART's large model (CLAP-ART(BEATs-RVQ)-large). 
For both CLAP-ART(BEATs-RVQ), we also report the metric scores of the model that achieved the best SPIDEr score out of six experiments as a reference.

The results showed that even with the base model of BART, CLAP-ART(BEATs-RVQ)-base outperformed all other methods, except for LOAE~\cite{liu2024enhancing}. 
The metric scores of CLAP-ART(BEATs-RVQ)-base were averaged over six experiments with different random seeds, demonstrating the superior performance of CLAP-ART(BEATs-RVQ)-base regardless of random seed variations.
CLAP-ART(BEATs-RVQ)-large further improved the averaged metric scores.
The best-performing CLAP-ART(BEATs-RVQ) models among the six experiments achieved a SPIDEr score comparable to that of LOAE, an LLM-based method. 
These results suggest that the performance improvement based on our hypothesis in CLAP-ART is a significant contribution comparable to the use of LLM.

\section{Conclusions}
This paper proposed CLAP-ART, which improves AAC performance by utilizing ``semantic-rich and discrete'' tokens, which capture fine-grained semantic contexts of general sounds.
CLAP-ART is based on our hypothesis that semantic-rich discrete tokens are beneficial inputs for fine-tuning language models.
To obtain such discrete tokens, we designed ART by leveraging semantically rich AR.
We implemented CLAP-ART to take ART discrete tokens and CLAP audio embedding as input for BART.
Experiments on two AAC benchmarks showed that CLAP-ART outperformed the baseline EnCLAP, supporting our hypothesis.

Future work includes investigating the integration of LLM-based methodologies and post-processing strategies, such as reranking~\cite{Kim2024} and CLAP-based text decoding~\cite{chen2024slam}.

\clearpage
\section{Acknowledgements}
This work was partially supported by JST Strategic International Collaborative Research Program (SICORP), Grant Number JPMJSC2306, Japan.
\bibliographystyle{IEEEbib}
\bibliography{refs}

\begin{thebibliography}{10}

\bibitem{drossos2017automated}
K.~Drossos, S.~Adavanne, and T.~Virtanen,
\newblock ``Automated audio captioning with recurrent neural networks,''
\newblock in {\em Workshop Appl. Signal Process. Audio Acoust. (WASPAA)}. IEEE, 2017, pp. 374--378.

\bibitem{mei2022automated}
X.~Mei, X.~Liu, M.~D Plumbley, and W.~Wang,
\newblock ``Automated audio captioning: An overview of recent progress and new challenges,''
\newblock {\em EURASIP J. Audio Speech Music Process.}, vol. 2022, no. 1, pp. 26, 2022.

\bibitem{xu2024beyond}
X.~Xu, Z.~Xie, M.~Wu, and K.~Yu,
\newblock ``Beyond the status quo: A contemporary survey of advances and challenges in audio captioning,''
\newblock {\em IEEE/ACMTrans. Audio Speech Lang. Process.}, vol. 32, pp. 95--112, 2024.

\bibitem{koizumi2020audio}
Y.~Koizumi, Y.~Ohishi, D.~Niizumi, D.~Takeuchi, and Masahiro Y.,
\newblock ``Audio captioning using pre-trained large-scale language model guided by audio-based similar caption retrieval,''
\newblock {\em arXiv preprint arXiv:2012.07331}, 2020.

\bibitem{gontier2021automated}
F.~Gontier, R.~Serizel, and C.~Cerisara,
\newblock ``Automated audio captioning by fine-tuning bart with audioset tags,''
\newblock in {\em Proc. Workshop Detect. Classif. Acoust. Scenes Events (DCASE)}, 2021.

\bibitem{kim2024enclap}
J.~Kim, J.~Jung, J.~Lee, and S.~H. Woo,
\newblock ``Enclap: Combining neural audio codec and audio-text joint embedding for automated audio captioning,''
\newblock in {\em Proc. IEEE Int. Conf. Acoust. Speech Signal Process. (ICASSP)}, 2024, pp. 6735--6739.

\bibitem{mei2024wavcaps}
X.~Mei, C.~Meng, H.~Liu, Q.~Kong, T.~Ko, C.~Zhao, M.~D Plumbley, Y.~Zou, and W.~Wang,
\newblock ``Wavcaps: A chatgpt-assisted weakly-labelled audio captioning dataset for audio-language multimodal research,''
\newblock {\em IEEE/ACM Trans. Audio Speech Lang. Process.}, 2024.

\bibitem{wu2024improving}
S.-L. Wu, X.~Chang, G.~Wichern, J.~Jung, F.~Germain, J.~Le~Roux, and S.~Watanabe,
\newblock ``Improving audio captioning models with fine-grained audio features, text embedding supervision, and llm mix-up augmentation,''
\newblock in {\em Proc. IEEE Int. Conf. Acoust. Speech Signal Process. (ICASSP)}. IEEE, 2024, pp. 316--320.

\bibitem{ghosh2024recap}
S.~Ghosh, S.~Kumar, C.~K. Reddy~Evuru, R.~Duraiswami, and D.~Manocha,
\newblock ``Recap: Retrieval-augmented audio captioning,''
\newblock in {\em Proc. IEEE Int. Conf. Acoust. Speech Signal Process. (ICASSP)}, 2024, pp. 1161--1165.

\bibitem{haji2024taming}
M.~Haji-Ali, W.~Menapace, A.~Siarohin, G.~Balakrishnan, S.~Tulyakov, and V.~Ordonez,
\newblock ``Taming data and transformers for audio generation,''
\newblock {\em arXiv preprint arXiv:2406.19388}, 2024.

\bibitem{liu2024enhancing}
J.~Liu, G.~Li, J.~Zhang, H.~Dinkel, Y.~Wang, Z.~Yan, Y.~Wang, and B.~Wang,
\newblock ``Enhancing automated audio captioning via large language models with optimized audio encoding,''
\newblock in {\em Proc. Interspeech}, 2024.

\bibitem{chen2023beats}
S.~Chen, Y.~Wu, C.~Wang, S.~Liu, D.~Tompkins, Z.~Chen, W.~Che, X.~Yu, and F.~Wei,
\newblock ``{BEAT}s: Audio pre-training with acoustic tokenizers,''
\newblock in {\em Proc. Int. Conf. Mach. Learn. (ICML)}, Jul. 2023, pp. 5178--5193.

\bibitem{gemmeke2017audioset}
J.~F. Gemmeke, D.~P.~W. Ellis, D.~Freedman, A.~Jansen, W.~Lawrence, R.~C. Moore, M.~Plakal, and M.~Ritter,
\newblock ``{Audio Set}: An ontology and human-labeled dataset for audio events,''
\newblock in {\em Proc. IEEE Int. Conf. Acoust. Speech Signal Process. (ICASSP)}, 2017, pp. 776--780.

\bibitem{wu2023large}
Y.~Wu, K.~Chen, T.~Zhang, Y.~Hui, T.~Berg-Kirkpatrick, and S.~Dubnov,
\newblock ``Large-scale contrastive language-audio pretraining with feature fusion and keyword-to-caption augmentation,''
\newblock in {\em Proc. IEEE Int. Conf. Acoust. Speech Signal Process. (ICASSP)}, 2023, pp. 1--5.

\bibitem{defossez2022high}
A.~D{\'e}fossez, J.~Copet, G.~Synnaeve, and Y.~Adi,
\newblock ``High fidelity neural audio compression,''
\newblock {\em Trans. Mach. Learn. Res.}, 2023.

\bibitem{lewis2020bart}
Mike Lewis, Y.~Liu, N.~Goyal, M.~Ghazvininejad, A.~Mohamed, O.~Levy, V.~Stoyanov, and L.~Zettlemoyer,
\newblock ``{BART}: Denoising sequence-to-sequence pre-training for natural language generation, translation, and comprehension,''
\newblock in {\em Proc. 58th Annu. Meet. Assoc. Comput. Linguist. (ACL)}, 2020, pp. 7871--7880.

\bibitem{liu2024semanticodec}
H.~Liu, X.~Xu, Y.~Yuan, M.~Wu, W.~Wang, and M.~D Plumbley,
\newblock ``Semanticodec: An ultra low bitrate semantic audio codec for general sound,''
\newblock {\em arXiv preprint arXiv:2405.00233}, 2024.

\bibitem{zhang2024speechtokenizer}
X.~Zhang, D.~Zhang, S.~Li, Y.~Zhou, and X.~Qiu,
\newblock ``Speechtokenizer: Unified speech tokenizer for speech language models,''
\newblock in {\em Proc. Int. Conf. Learn. Represent.}, 2024.

\bibitem{hsu2021hubert}
W.-N. Hsu, B.~Bolte, Y.-H~H. Tsai, K.~Lakhotia, R.~Salakhutdinov, and A.~Mohamed,
\newblock ``Hubert: Self-supervised speech representation learning by masked prediction of hidden units,''
\newblock {\em IEEE/ACM Trans. Audio Speech Lang. Process.}, vol. 29, pp. 3451--3460, 2021.

\bibitem{zeghidour2021soundstream}
Neil Zeghidour, Alejandro Luebs, Ahmed Omran, Jan Skoglund, and Marco Tagliasacchi,
\newblock ``{SoundStream}: An end-to-end neural audio codec,''
\newblock {\em IEEE/ACM Trans. Audio Speech Lang. Process.}, vol. 30, pp. 495--507, 2022.

\bibitem{Baevski2020vq-wav2vec}
Alexei Baevski, Steffen Schneider, and Michael Auli,
\newblock ``vq-wav2vec: Self-supervised learning of discrete speech representations,''
\newblock in {\em International Conference on Learning Representations}, 2020.

\bibitem{mousavi2024how}
P.~Mousavi, J.~Duret, S.~Zaiem, L.~{Della Libera}, A.~Ploujnikov, C.~Subakan, and M.~Ravanelli,
\newblock ``How should we extract discrete audio tokens from self-supervised models?,''
\newblock in {\em Interspeech 2024}, 2024, pp. 2554--2558.

\bibitem{kim2019audiocaps}
C.~D. Kim, B.~Kim, H.~Lee, and G.~Kim,
\newblock ``Audiocaps: Generating captions for audios in the wild,''
\newblock in {\em Proc. Conf. N. Am. Chapter Assoc. Comput. Linguist. (NAACL)}, 2019, pp. 119--132.

\bibitem{drossos2020clotho}
K.~Drossos, S.~Lipping, and T.~Virtanen,
\newblock ``Clotho: An audio captioning dataset,''
\newblock in {\em Proc. IEEE Int. Conf. Acoust. Speech Signal Process. (ICASSP)}. IEEE, 2020, pp. 736--740.

\bibitem{loshchilov2017decoupled}
I.~Loshchilov and F.~Hutter,
\newblock ``Decoupled weight decay regularization,''
\newblock {\em arXiv preprint arXiv:1711.05101}, 2017.

\bibitem{denkowski2014meteor}
M.~Denkowski and A.~Lavie,
\newblock ``Meteor universal: Language specific translation evaluation for any target language,''
\newblock in {\em Proc. Workshop Stat. Mach. Transl.}, June 2014, pp. 376--380.

\bibitem{vedantam2015cider}
R.~Vedantam, C.~Lawrence~Zitnick, and D.~Parikh,
\newblock ``{CIDEr: Consensus-based image description evaluation},''
\newblock in {\em IEEE Conf. Comput. Vis. Pattern Recognit. (CVPR)}, 2015, pp. 4566--4575.

\bibitem{anderson2016spice}
P.~Anderson, B.~Fernando, M.~Johnson, and S.~Gould,
\newblock ``{SPICE: Semantic propositional image caption evaluation},''
\newblock in {\em Eur. Conf. Comput. Vis. (ECCV)}, 2016, pp. 382--398.

\bibitem{liu2017improved}
S.~Liu, Z.~Zhu, N.~Ye, S.~Guadarrama, and K.~Murphy,
\newblock ``Improved image captioning via policy gradient optimization of spider,''
\newblock in {\em IEEE Int. Conf. Comput.Vis. (ICCV)}, 2017, pp. 873--881.

\bibitem{zhou2022can}
Z.~Zhou, Z.~Zhang, X.~Xu, Z.~Xie, M.~Wu, and K.~Q Zhu,
\newblock ``Can audio captions be evaluated with image caption metrics?,''
\newblock in {\em Proc. IEEE Int. Conf. Acoust. Speech Signal Process. (ICASSP)}. IEEE, 2022, pp. 981--985.

\bibitem{Kim2024}
J.~Kim, M.~Jeon, J.~Jung, S.~H. Woo, and J.~Lee,
\newblock ``Enclap++: Analyzing the enclap framework for optimizing automated audio captioning performance,''
\newblock in {\em Proc. Workshop Detect. Classif. Acoust. Scenes Events (DCASE),}, 2024, pp. 61--65.

\bibitem{chen2024slam}
W.~Chen, Z.~Ma, X.~Li, X.~Xu, Y.~Liang, Z.~Zheng, K.~Yu, and X.~Chen,
\newblock ``Slam-aac: Enhancing audio captioning with paraphrasing augmentation and clap-refine through llms,''
\newblock in {\em Proc. IEEE Int. Conf. Acoust. Speech Signal Process. (ICASSP)}, 2025.

\end{thebibliography}

\end{document}